\documentclass[prb,twocolumn,superscriptaddress,aps]{revtex4-1}
\usepackage{amsmath}
\usepackage{color}
\usepackage{dcolumn}
\usepackage{bm}
\usepackage{slashed}
\usepackage[toc,page]{appendix}
\usepackage{float}
\usepackage{amssymb}
\usepackage{graphicx}

\def\gapp{\lower.35em\hbox{$\stackrel{\textstyle>}{\sim}$}}
\def\lapp{\lower.35em\hbox{$\stackrel{\textstyle<}{\sim}$}}

\begin{document}
\bibliographystyle{apsrev4-1}
%

\title{Finite frequency magnetoelectric response of three dimensional Topological Insulators}

\author{Adolfo G. Grushin} 
\affiliation{Instituto de Ciencia de Materiales de Madrid, CSIC, Cantoblanco, E-28049 Madrid,
Spain.}
\author{Fernando de Juan} 
\affiliation{Department of Physics, Indiana University, Bloomington, IN 47405, USA}
\date{\today}

\begin{abstract}
Topological insulators with a time reversal symmetry breaking perturbation near the surface present
a magnetoelectric response that is quantized when the frequency of the probing fields is much
smaller than the surface gap induced by the perturbation. In this work we describe the intrinsic
finite frequency magnetoelectric response of topological insulators for frequencies of the order and
larger than the
surface gap, including the experimentally relevant case where the system is metallic. This response
affects physical observable quantities and will give rise to new finite frequency phenomena of
intrinsic topological origin.
\end{abstract}

\maketitle

\section{Introduction}

It is well understood \cite{LLv8} that under a time dependent external electromagnetic perturbation
any given material will develop a time dependent response, whenever the characteristic frequencies
of the perturbation are larger than the characteristic frequencies that induce a polarization or a
magnetization in the material. The linear response of conventional dielectrics is characterized by a
dielectric function $\varepsilon(\omega)$ and a magnetic permeability $\mu(\omega)$. There is in
fact a wider class of materials, known as magnetoelectric materials \cite{EMS06}, which effectively
introduce other response functions that couple electric and magnetic fields and which in general can
also depend on the frequency. Perhaps the most striking case of the latter are the recently
discovered three-dimensional topological insulators (TI) \cite{RevZhang11,RevHas11}. They have been
predicted to host a quantized magnetoelectric term in the action, topological in origin
\cite{QHZ08,EMV09} of the form $S=\int dtd^3 x(\alpha/4\pi^2)\theta\mathbf{E}\cdot\mathbf{B}$, where
$\alpha=e^2/\hbar c$ and $\theta = \pi$, only physically relevant when time reversal symmetry is
broken which implies that the surface states that characterize these materials are gapped. Despite the fact that
this term remains experimentally elusive, there has been much ongoing work on its consequences. It
has has been predicted to give rise to a plethora of phenomena including the Kerr and Faraday
rotation of light determined by the fine structure constant \cite{MQDZ10,TM10} and a repulsive
Casimir effect \cite{GC11,GRC11}, where the region of repulsion is determined by $\theta$.
Physically, these approaches are valid only when the frequencies of the relevant fields are much
smaller than the surface gap $m$, and the topological magnetoelectric term $\theta$ is independent
of the frequency and quantized. In general, the magnetoelectric response will depend on the
frequency, and permeate into physical observables, just as the dielectric function or the magnetic
permeability do, modifying all the described phenomena related to this topological term.\\
In this work, we derive the finite frequency magnetoelectric response of a model hamiltonian which
captures the basic features of a three dimensional TI. To do so, we will generalize the method
introduced in 
Ref. \onlinecite{QHZ08} to finite frequency, relating the response of TIs to that of an effective model with an
extra 
dimension that behaves as a higher dimensional analogue of the Quantum Hall Effect (QHE)
\cite{ZH01}. 
This approach has been shown to be helpful to understand the topological origin of this response,
and it is also an
efficient 
computational tool in practice. It has proven useful to predict the magnetoelectric response in a
related 
physical situation \cite{BQ11,Ber11,BR12} where the bulk of the TI is assumed to be doped. 
In this particular case, the magnetoelectric response is not quantized if the chemical potential is
outside
the band gap, a behaviour that may be interpreted as arising from the corresponding anomalous QHE
analogue in five dimensions.\\
Here we extend these analysis to a more general and potentially relevant experimental
situation where both the frequency and the chemical potential are kept finite, a case that can also
be understood as descending from a five dimensional finite frequency QHE at
finite chemical potential. As a consistency check we will show how known results are recovered in
the appropriate limits, giving further physical insight into them.
\section{The model}
Consider the lattice hamiltonian introduced in Ref. \onlinecite{QHZ08}, which captures the low
energy description of a generic TI, for instance Bi$_{2}$Se$_{3}$ \cite{LQZ10,Ber11}. This
hamiltonian can be written as $H = H_0 + H_M$ with
\begin{align}\label{H}
H_0&=t\sum_{\bf{x}, \mathbf{s}} c^{\dagger}_{\mathbf{x}}\frac{\Gamma_0 - i\Gamma_s}{2}
c_{\mathbf{x}+ \hat{\mathbf{s}}} + h.c.
-3t  c^{\dagger}_{\mathbf{x}} \Gamma^0 c_{\mathbf{x}}, \\
\label{HM1}
H_M &= M \sum_{\mathbf{x}}c^{\dagger}_{\mathbf{x}} \Gamma^0 c_{\mathbf{x}},
\end{align}
with $\bf{x}$ running through all unit cells, $s=1,2,3$, and $\bf\hat s$ is the lattice vector in
the
$s$ direction. $\Gamma_{\mu}$ are defined as the set of $4\times4$ matrices that
satisfy $\lbrace\Gamma_{\mu},\Gamma_{\nu}\rbrace=2\delta_{\mu\nu},$ with
$\mu,\nu=0,1,2,3,4$ including an extra $\Gamma_4$ that will be used shortly. In momentum space this
can
be written as
\begin{equation}\label{Ha}
H(\mathbf{k})=  \left(t \sum_{s=1}^{3} \cos(k_s) +M -3t\right)\Gamma_0 + t \sum_{s=1}^{3} \sin(k_s)
\Gamma_s 
\end{equation}
These models can be thought of as lattice models
that host an
odd number of low energy massive Dirac fermions \footnote{A term of the type
$\epsilon(\mathbf{k})\mathrm{I}_{4\times4}$ can also be
included but it does not affect the topological properties so we neglect it for simplicity}. For
example, for  $|M|<t$ there is a single Dirac fermion at $\mathbf{k}=0$ of gap $M$. \\
The presence of a boundary in the hamiltonian can be modeled by making the gap position dependent,
promoting \eqref{HM1} to
\begin{equation}\label{HM2}
H_M = \sum_{\mathbf{x}} M\cos\theta(\mathbf{x}) c^{\dagger}_{\mathbf{x}} \Gamma_{0} c_\mathbf{x}.
\end{equation}
 This accounts for the band inversion by setting
$\theta(-\infty)=\pi$ in the bulk of the
TI and $\theta(\infty)=0$. The specific dependence of $\theta$ on $x$ will not be needed for our
purposes, only its
asymptotic values. Both experimentally \cite{X09} and from \emph{ab initio} calculations
\cite{LQZ10} the bulk band gap is well approximated by $M=0.3$ eV for Bi$_{2}$Se$_{3}$. A time
reversal symmetry breaking perturbation may generically be included as\cite{BQ11}
\begin{equation}\label{Hm}
H_m = \sum_{\mathbf{x}} m\sin\theta(\mathbf{x}) c^{\dagger}_{\mathbf{x}} \Gamma_{4} c_{\mathbf{x}}, 
\end{equation}
which is localized at the boundary and opens a surface gap $m$. This surface gap can arise from
doping the TI with magnetic impurities, and has been measured to be $m \sim 50$ meV
\cite{CCA10,WXX10}.

\section{Finite frequency electromagnetic response of a topological insulator}

To obtain the finite frequency response of a TI system to electromagnetic fields in the
presence of
$\theta(x)$, we will first generalize the original procedure devised in Ref. \onlinecite{QHZ08} to
finite frequency. In what follows, we compute the current response of the system with a generalized
Kubo formula in a way that the effect of $\theta(x)$ is included in a manifestly perturbative
fashion along the derivation. Our starting point is to consider the current density at some
particular point in space-time $x_0$:
\begin{equation}
j^{\mu}(x_0) = \frac{\delta S}{ \delta A_{\mu}(x_0)},
\end{equation}
where $S$ is the action functional of the system. For a profile $\theta(x)$ that is smooth over
length scales $l_m\equiv 1/(v_F m)$ (this is,
$|\vec\nabla \theta| << 1/l_m $), the current at $x_0$ is mainly determined by $\theta$ around
$\theta(x_0) \equiv
\theta_0$, because correlation functions decay exponentially with $l_m$. We may therefore
include its effects in perturbation theory in $\partial_i\theta$, which is by assumption small. For
the calculation of
$j^{\mu}(x_0)$,  we thus approximate\cite{BQ11}:
\begin{equation}
\theta(x) \approx \theta(x_0) + \left. 
\partial_{i} \theta\right|_{x=x_0} (x^{i}-x^{i}_{0}) +\cdots,
\end{equation}
in the hamiltonian $H=H_{0}+H_{M}+H_{m}$ 
defined by \eqref{H},\eqref{HM2} and \eqref{Hm} respectively. To first order in $\partial_{i}
\theta$ 
the mass terms read
\begin{align}\label{Ham}
H_M& + H_m = \sum_{\mathbf{x}} c^{\dagger}_{\mathbf{x}} (M\cos\theta_0 \Gamma_{0}+m\sin\theta_0
\Gamma_{4}) 
c_{\mathbf{x}}
\nonumber\\
+& \left. \partial_{i} \theta \right|_{\mathbf{x}_0}\sum_{\mathbf{x}} (x^{i}-x^{i}_{0}) 
c^{\dagger}_{\mathbf{x}}
(-M\sin\theta_0 \Gamma_{0}+m\cos\theta_0 \Gamma_{4})
  c_{\mathbf{x}}.
\end{align} 
Note that in this hamiltonian $\theta_0$ is just a constant parameter.
\\
We can now compute the current response at $x_0$ when a time dependent uniform electric field is
applied to the system. This is done by computing the expectation value of $j^{\mu}$ to first order 
in both $\partial_i \theta$ and the electromagnetic field $A_{\mu}$, with a generalized Kubo
formula
\begin{equation}
j^{i}(x_0) = \left. \partial_{s}  \theta \right|_{\mathbf{x}_0} \sum_{ x, x'} \left< 
\hat{J}^i(x_{0}) \hat{J}^j(x) \hat{J}_{\theta}^{s}(x') \right> A_j(x)
\end{equation}
where $i,j,s=1,2,3$, repeated indices summation is implied and $x_{0},x,x'$ are full space-time
variables. In this expression $\hat{J}^{i}(x)$ are the current operators, and
$\hat{J}_{\theta}^{s}(x)$ is the operator attached to $\partial_s \theta$ in \eqref{Ham} which
defines the following vertex in momentum space 
\begin{equation}\label{vertexf}
J_{\theta}^s(k)= (-M\sin\theta_0 \Gamma_{0}+m\cos\theta_0 \Gamma_{4})\partial_{k_s} \equiv
J_{\theta} \partial_{k_s}.
\end{equation}
With this, the Fourier transform of the current reads
\begin{align}\label{curr}
&j^{i}(x_0) = \left. \partial_{s}  \theta \right|_{\mathbf{x}_0} \int_{BZ}  \dfrac{d^4 p}{(2\pi)^4}
e^{-i p x_0} A^{j}_p  \int_{BZ} \dfrac{d^4 k}{(2\pi)^4} \\\nonumber
& \times \Big[ \mathrm{Tr} J^i_{k-p/2} G_{k-p} J^j_{k-p/2}
G_k J_{\theta} \partial_{k_s} G_k 
+ \left\lbrace 
\begin{array}{c}
p \hspace{1mm} \longleftrightarrow  -p\\
i \longleftrightarrow  j
\end{array}\right\rbrace \Big] .
\end{align}
where the integral spans the entire Brillouin Zone (BZ). The electronic Green's function, which
depends on a four-momentum vector
$k=(k_{0},\mathbf{k})$ is given
by $G(k,\theta_{0})=(k_{0}-H(\mathbf{k},\theta_{0}))^{-1}$ defined through the Fourier transformed
hamiltonian
\begin{align}\label{Ham2}
H(\mathbf{k},\theta_{0}) &= H(\mathbf{k})+(M\cos\theta_0 \Gamma_{0}+m\sin\theta_0 \Gamma_{4}) , 
\end{align}
that depends parametrically on $\theta_{0}$, the bulk mass $M$ and the surface mass $m$. The current
vertices are defined as $J^{i}(k)=\frac{\partial H(\mathbf{k},\theta_{0})}{\partial k_{i}}$,
$J^{0}=\mathrm{I}_{4\times 4}$  with $i=1,2,3$. In the derivation we have omitted terms arising
from vertices with higher derivatives of $H(\mathbf{k},\theta_{0})$ with respect to
$k$\cite{Bas09}, that will not contribute to the magnetoelectric response.  \\
Before we proceed further, it is worth noting that this equation may also be written as
\begin{equation}
j^{i}(x_0)  = \left. \partial_{s}  \theta \right|_{\mathbf{x}_0} \int_{BZ}  \dfrac{d^4 p}{(2\pi)^4}
e^{-i p x_0} \partial_{q_s} \left[\Pi^{ij}_{4}(p,q,\theta_{0})\right]_{q=0} A^{j}_p 
\end{equation}
where
\begin{align}\label{Pi4}
&\Pi^{\mu\nu}_{4}(p,q,\theta_{0})=-ie^2\int_{BZ} \dfrac{d^4
k}{(2\pi)^4} \\ 
&\times \Big[ \mathrm{Tr} J^i_{k-(p+q)/2} G_{k-p} J^j_{k-p/2}
G_k J_{\theta} G_{k-q} 
+ \left\lbrace 
\begin{array}{c}
p \hspace{1mm} \longleftrightarrow  -p+q\\
i \longleftrightarrow  j
\end{array}\right\rbrace \Big] .\nonumber
\end{align}
and again the identity holds disregarding higher derivatives of $H(\mathbf{k},\theta_{0})$. $\Pi^{ij}_4(p,q)$ can be
considered the response function to $\delta \theta = \theta_0 - \theta(x)$
\begin{equation}
j^{i}(x_0)  = \int_{BZ} \dfrac{d^4 p}{(2\pi)^4}\dfrac{d^4 q}{(2\pi)^4}
e^{-i (p+q) x_0} \Pi^{ij}_{4}(p,q,\theta_{0}) A^{j}_p \delta \theta_q ,
\end{equation}
which is the generalization to finite frequency and momenta of Ref. \onlinecite{QHZ08}. Equation
\eqref{curr} represents an equivalent statement that features an
explicit small parameter throughout the derivation. \\
Consider now the case where the boundary of the TI is in the $z$ direction, so that
$\theta(\mathbf{x})=\theta(z)$.
A uniform but time dependent electric field $E_{j}$ in momentum space can be written in terms of an 
external vector potential $A_i$ that is constant in space, so that
$A_{i}(p_0,\mathbf{p}) = \delta(\mathbf{p}) A_{i}(p_0)$. \\
The total current density in the
$xy$ plane, shown to be quantized in the DC limit \cite{QHZ08,BQ11}
is defined as $\mathcal{J}_{2D}^{i}=\int dz j^{i}(z)$ with $i=x,y$. The finite frequency
generalization of this quantity,
i.e. the integrated current density, thus reads 
\begin{equation}
\mathcal{J}^{i}_{2D}(p_0) = \int dz_0 \partial_{z_{0}}\theta(z_0)
\partial_{q_z} \left[\Pi^{ij}_{4}(p,q,\theta_{0})\right]_{q=0,\mathbf{p}=0} A^{j}_{p_0} .
\end{equation}
With the change of variables $\int_{-\infty} ^{\infty} dz_0  \partial_{z_0}\theta =
\int_{-\pi}^{0} d\theta$ the current is finally
\begin{eqnarray}\label{freqdepcurr4}
\mathcal{J}^{i}_{2D}(p_{0})&=& 2\pi\int_{-\pi}^0 \dfrac{d \theta}{2\pi} \partial_{q_z}
\left[\Pi^{ij}_{4}(p,q,\theta)\right]_{q=0,\mathbf{p}=0} A^{j}_{p_0}.
\end{eqnarray}
As in the static case, the parameter $\theta$ can be thought of as the fifth coordinate of a 4+1
model described by $H(\mathbf{k},\theta)$, whose response functions are integrated only over half of
the
BZ, because $0 \le \theta \le \pi$. The topological finite frequency response of a 3D TI is thus
intimately related to the finite frequency response of a $D=4+1$ insulator.\\
Consider now the experimentally relevant limit where $m\ll M$. In this limit the response in
\eqref{freqdepcurr4} can be obtained analitically. This is due to the fact that the low energy
physics of
hamiltonian (\ref{Ham2}), considered now as $4+1$ hamiltonian in half of the BZ, is dominated by an
effective 4+1 Dirac fermion of gap $m$ located at $(\mathbf{k},\theta)=(0,\pi/2)$, where the $4+1$
analogue
of the Berry curvature is largest. Integrals in the five
dimensional BZ are thus well approximated by a
region of momenta around $(0,\pi/2)$ within some cut-off $\Lambda$, and the effective
hamiltonian in the vicinity of that point is given by
\begin{align}\label{expand}
H(\mathbf{k},\theta) \approx &H(0,\pi/2) + \left.\partial_{k_i} H\right|_{(0,\pi/2)} k_i +
\left.\partial_{\theta} H\right|_{(0,\pi/2)} \tilde{\theta} \nonumber \\
= & \Gamma^i k_i + M  \tilde{\theta}\Gamma^0 + m \Gamma^5
\end{align}
with $\theta = \pi/2 + \tilde{\theta}$. We can identify this hamiltonian as that of a 4+1 Dirac
fermion where $k_4 = M\tilde{\theta}$. The cut-off for this model is of order $\Lambda \approx
M$, and for $\omega<<M$ may be taken to infinity. Within this approximation, the $\theta$ integral
in \eqref{freqdepcurr4} is
\begin{align}
&\Pi_5^{ij0}(p,q) \equiv  \int_{-\pi}^0 \dfrac{d \theta}{2\pi}  \Pi^{ij}_{4}(p,q,\theta) \approx \frac{-ie^2}{M} 
\int^{\infty} \frac{d^5 k}{(2\pi)^5} \nonumber \\ 
&\mathrm{Tr} \Big[\Gamma^i G_{k-p}
\Gamma^j G_{k}M \Gamma^0 G_{k-q} +\Gamma^j G_{k+p-q}
\Gamma^i G_{k}M \Gamma^0 G_{k-q}\Big],\label{FD5}
\end{align}
where we have used $d \tilde{\theta} = dk_4/M$, and eq. \eqref{Pi4} with the current vertices
approximated
around $(0,\pi/2)$: $J^{i}=\Gamma^i$, $J^{\theta}= M \Gamma^0$. The Green functions in these
expressions are those of a 4+1 Dirac fermion, obtained as $G(k)=(k_0 -H(\mathbf{k}))^{-1}$ 
from eq. \eqref{expand}, where now $\mathbf{k}$ has four components.
Consequently, the function $\Pi_5^{ij0}$ corresponds to the Dirac fermion triangle diagrams shown in
Fig. \ref{FD} that can be computed analytically with standard methods that are detailed in appendix
A (see also Ref. \onlinecite{DW94}), and can be considered as the
optical response of the five dimensional analogue of the QHE.
\begin{figure}
\includegraphics[scale=0.6]{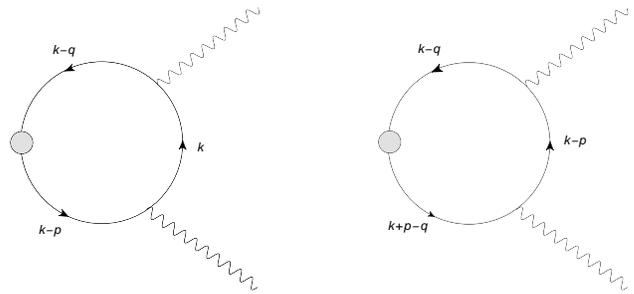}
\caption{\label{FD}Feynman diagrams corresponding to eq. \eqref{FD5}. The second diagram is obtained
from the first by by $i \leftrightarrow j$ and $p \leftrightarrow -p +q $. The grey dot represents the $\partial_{s}\theta$ vertex.}
\end{figure}
The magnetoelectric response is given by the antisymmetric part of the diagram and thus the total
current finally reads
\begin{eqnarray}
\nonumber
\mathcal{J}^{i}_{2D}(p_{0})&=& 2\pi\Pi_{5}(p_{0},\mu)\epsilon^{ij}E_{j}(p_{0}) \\
\label{sigma}
&\equiv & \sigma(p_{0},\mu)\epsilon^{ij}E_{j}(p_{0}),
\end{eqnarray}
where the function 
\begin{equation}\label{Pi5final}
 \Pi_{5}(p,q) = \frac{\epsilon_{ij}}{2} \frac{1}{p_0} \partial_{q_z}
\left[\Pi^{ij0}_{5}(p,q)\right]_{q=0,\mathbf{p}=0},
\end{equation}
and we have used that $E^{j}_{p}=i p_{0}A^{j}_{p}$. \\
Equations \eqref{sigma} and \eqref{Pi5final} define the finite frequency response
of a TI, which we proceed to evaluate in the next section for different experimentally relevant
scenarios.

\section{Results}
\begin{figure}
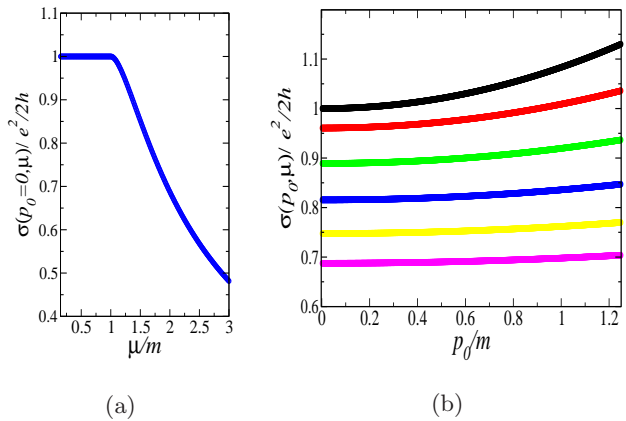

\begin{minipage}{.39\linewidth}
\begin{center}
\begin{center}
\includegraphics[width=30mm,height=47mm]{Pivsmu}
\end{center}
\begin{center}
(a) 
\end{center}
\end{center}
\end{minipage}
\begin{minipage}{.59\linewidth}
\begin{center}
\includegraphics[width=47mm,height=47mm]{Pi5vsmup0}
\begin{center}
(b) 
\end{center}
\end{center}
\end{minipage}
\caption{\label{Fig:Pi5vsmu} $\sigma(p_{0},\mu)$ as a function of (a) the chemical potential $\mu$
for zero frequency in units of the surface gap $m$ and (b) as a function of the external frequency
$p_{0}$ for $\mu/m=1.0-2.0$ in steps of $0.2$ (top to bottom). There is a quantization plateau
whenever $|\mu| \leq m$ and a decay for $\mu\geq m$ given by (\ref{zerofreq}). For finite
frequencies and whenever $|\mu|\geq m$ is satisfied the DC value is not quantized and given by
\eqref{mup0}.}
\end{figure}

In this section we compute the function $\sigma(p_{0},\mu)$ defined above which determines the 
response of the TI to an external electromagnetic field of finite frequency.\\
Firstly, it is possible to restore the dependence on the chemical potential since 
nothing in the above argument depends on whether or not the chemical potential is finite as long as
$\mu <<M$. For a massive Dirac fermion at zero chemical potential, the function
$\Pi_{5}(p_{0},\mu=0)$ can be analytically computed. 
The final analytical expression to which one arrives depends only on the surface gap $m$ 
and is given by (we refer the reader to appendix A for details)
\begin{equation}\label{firstpeak}
\sigma(p_{0},\mu=0)=\dfrac{1}{2}\dfrac{e^2}{h}\dfrac{m}{p_{0}}\log\Big
|\dfrac{2m+p_{0}}{2m-p_{0}}\Big |.
\end{equation} 
\begin{figure}
\begin{center}
\includegraphics[scale=0.25]{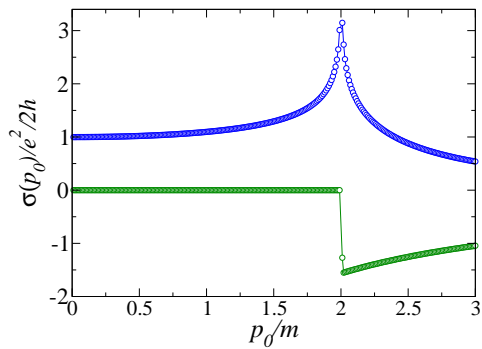}
\end{center}
\caption{\label{Fig:4+1plot} Real (top) and imaginary (bottom) parts of $\sigma(p_{0},\mu=0)$ given
by \eqref{firstpeak} as a function of $p_{0}$ in units of the surface gap $m$. The quantization is
lost at higher frequency and a logarithmic singularity appears at $p_{0}=2m$}
\end{figure} 
This function, plotted in Fig. \ref{Fig:4+1plot}, governs the finite frequency response of a TI and
is one of the central results of this work. The quantization of the real part at
low frequencies is broken down at finite frequency giving rise to a logarithmic divergence at $p_{0}
= 2m$. Therefore, close to this range of frequencies, the $\theta$ term will dominate the
electromagnetic response of the TI. This is particularly important for Casimir type experiments
\cite{GC11,GRC11}, where the interplay between the optical properties of ordinary and topological
response determines not only the sign of the force, but also at what distance does the crossover
between attractive and repulsive behavior happens. The precise way this response alters the Casimir
force is an interesting issue on its own, and it is left for a future study.\\
It is important to note that the analytic result \eqref{firstpeak} coincides exactly with the
optical Hall conductivity of a massive $D=2+1$ dimensional Dirac fermion. The DC response of a
single Dirac fermion is quantized to $e^2/2h$, which is consistent with the fact that in the lattice
model the integrals span only half of the BZ. We thus recover the well known result that the
boundary of a TI with broken time reversal symmetry hosts a half-integer quantum Hall effect. \\
Being precise, this result should
not be interpreted as if there is a massive $D=2+1$ Dirac fermion somewhere in the system. Instead,
these results imply that the three-dimensional optical response of a TI is characterized by spatial
average in the $z$ direction of all the $\sigma_{xy}(z)$ Hall conductivities that occur wherever
there is a non zero gradient of $\partial_{z}\theta(z)$. This situation is relevant for the recent
experiments
described in \cite{CCA10,WXX10} where TI are doped with magnetic impurities that break time reversal
symmetry. It is remarkable nevertheless that a full $D=3+1$ calculation reduces to a $D=2+1$ result.
As will
be shown immediately below, this statement does not hold for the case of finite chemical
potential.\\
To do so, one should compute $\Pi_{5}(p_{0},\mu)$. This can be exactly evaluated for some cases
which we proceed to describe (technical details are left for appendix B).\\
One of them is the DC response at finite $\mu$. This limit was discussed earlier in Ref.
\cite{BQ11,Ber11,BR12} having obvious interest on its own since TI appear naturally doped in
experiments. The numerical evaluation for $\Pi_{5}(p_{0}\rightarrow 0,\mu)$ in the DC limit is
exactly given for our model by the expression:
\begin{equation}\label{zerofreq}
\sigma(p_{0}\rightarrow 0,\mu)=
\dfrac{e^2}{h}\left\{
	\begin{array}{ll}
		\dfrac{1}{2}\mathrm{sgn}(m)  & \mbox{if } |\mu | \leq m \\
		\dfrac{1}{4}\left[\dfrac{3m}{|\mu |}-\dfrac{m^3}{|\mu |^3} \right]& \mbox{if }  |\mu
| \geq m
	\end{array}
\right.
\end{equation} 
The result is shown in Fig. \ref{Fig:Pi5vsmu}(a). The
analytical expression reveals that although there is a quantized value at values of $|\mu|\leq m$
which also occurs for $D=2+1$ fermions \cite{KM05,CGV10}, already one can notice very important
differences with respect to the $D=2+1$ case. In $D=2+1$ it can be shown \cite{KM05,CGV10} that only
a term $m/|\mu|$ arises at fillings larger than the gap. In the present case however, there is a
second term which has a different behaviour and scales like $m^3/|\mu|^3$. This term is therefore
intrinsically related to the $D=3+1$ nature of the carriers which is only fully transparent in the
analytic result. The extra term turns the kink between the two regimes 
$|\mu | \leq m$ and $|\mu | \geq m$ smoother, making the curve in Fig. \ref{Fig:Pi5vsmu}(a)
differentiable at all $\mu$, in contrast with the $2+1$ result.\\
Finally it is possible to gain analytic insight into the regime where both the frequency $p_{0}$ and
$\mu$ are kept finite, a situation that can be clearly relevant for experimentally realistic
situations. It is easy to see that whenever $|\mu|\leq m$ and $p_0<2m$ the result is the same as in
\eqref{firstpeak}. However there is an experimentally more relevant situation when $|\mu|\geq m$ but
still $p_0<2m$. This is the case of a doped TI at finite frequency. Evaluating
$\Pi_{5}(p_{0}<<2m,|\mu|>m)$ one obtains the first non trivial order in $p_{0}$:
\begin{equation}\label{mup0}
\sigma(p_{0},|\mu|>m)=\dfrac{1}{4}\dfrac{e^2}{h}\left[ \dfrac{3m}{|\mu |}-\dfrac{m^3}{|\mu |^3}
+\dfrac{m p_{0}^2}{6|\mu |^3}\right],
\end{equation}
which is plotted as a function of the external frequency $p_{0}$ for different values of the
chemical potential $\mu$ in Fig. \ref{Fig:Pi5vsmu}(b). In the limit where $\mu=m$ this coincides
with the expansion of \eqref{firstpeak} when $p_{0}<<2m$. In this general case, the
quantization of the zero frequency value is also absent for finite values of $\mu$ and $p_{0}$. Thus
our results imply that the Kerr and Faraday rotation \cite{MQDZ10,TM10} will turn not to be
quantized in units of the fine structure constant if the samples are doped and/or if the frequency
of the probe is of the order of the surface gap, a common situation in actual experiments.
\section{Discussion and conclusions}
To summarize, the present findings will inevitably permeate into physically observable quantities
whenever optical probes have a frequency comparable or larger than the surface
gap. Given the sizes of the gaps, which are as large as $0.3$ eV for the bulk gap 
and $50$ meV for the surface gap \cite{CCA10,WXX10}, it should be possible to observe these effects
with infra-red probes, which are fully controllable within current state of the art technology
\cite{SBQ12}. The interplay between both scales can be studied in full lattice models and will be
the aim of a subsequent publication. \\
More elaborate scenarios, such as the proposed repulsive Casimir effect \cite{GC11,GRC11}, the
Topological Kerr and Faraday effect \cite{MQDZ10,TM10} or even the optical-modulator device proposed
in \cite{LWQ10} should be revisited. These findings can be generalized to other classes of
topological
materials such as certain classes of Weyl semi-metals that host a Carroll-Field-Jackiw term
\cite{CFJ90,Bal11,Gru12,ZB12} and also to higher dimensional
analogues of TI.\\
In conclusion we have calculated, the electrodynamic response of
TI at finite frequencies and finite chemical potential relating it to the response of a higher
dimensional analogue of the anomalous QHE. Beyond reasonable doubt,
these findings will permeate and strongly affect physical observables, just as any other finite
frequency response function. We have shown that there is a well defined action in this case and that
it is possible to define a quantity, the total current, which is not sensitive to the particular
time reversal breaking profile inside the TI but it is still dependent on the external frequency
characterizing the electromagnetic perturbation. These results pave the way to the understanding of
topological phenomena at finite frequency, which are bound to be relevant in current experimental
set-ups.\\

\section*{Acknowledgements}

A.G.G. is grateful to A. Cortijo and Shoucheng Zhang for discussions in the early stages of this
work. We thank M. A. H. Vozmediano, H. Ochoa, E. Cappelluti for useful insights. A.G.G
acknowledges support from spanish FIS2008-00124, FIS2011-23713 and PIB2010BZ-00512 (Brazil).
F. J. acknowledges funding under NSF Grant No. DMR-1005035. 

%

\begin{widetext}
\appendix

\section{Calculation of $\Pi_{5}(p_{0},\mu=0)$}

As described in the main text the main text $\Pi_{5}(p_{0},\mu)$ determines the finite frequency
response of the TI system. In this appendix we provide an alternative derivation starting from the
response of a $D=4+1$ system
and give details of how to compute it for $\mu=0$ leaving the $\mu\neq 0$ for the last appendix.\\
As shown in Ref. \onlinecite{QHZ08} the quantized DC magnetoelectric response of TI system can be
described as descending from a five dimensional analogue of the quantized integer quantum Hall
effect
(IQHE) \cite{ZH01}. Under this perspective the work of Refs. \onlinecite{BQ11,Ber11,BR12} for finite
chemical potential can be understood as arising from the corresponding anomalous QHE analogue in
five dimensions. Similarly, it is possible to reinterpret our results presented in the main text
as descending from a five dimensional finite frequency QHE at
finite chemical potential.\\
To describe the finite frequency response of the $D=4+1$ at $\mu=0$ we
couple the model to an external electromagnetic field $A_{\mu}$. After integrating out fermions, we
obtain an effective action for the gauge field. This effective action will generate an analogue of
the QHE described by a Chern-Simons like term which in momentum space reads:
\begin{equation}\label{4+1action}
S^{eff}_{4+1}=\int_{BZ} \dfrac{d^5q}{(2\pi)^5}\int
\dfrac{d^5p}{(2\pi)^5}\hspace{2mm}\Pi_{5}(p,q)\epsilon^{\mu\nu\rho\sigma\tau}A_{\mu}p_{\nu}A_{\rho}
q_{\sigma}A_{\tau},
\end{equation}
where $\epsilon^{\mu\nu\rho\sigma}$ is the Levi-Civita totally antisymmetric tensor and $A_{\mu}$ is
the electromagnetic gauge field. In real space, the $p_{\mu},q_{\nu}$ momenta turn into derivatives
and one recovers an action of the form $A\partial A\partial A$, which is a five dimensional analogue
of the IQHE action in $D=2+1$ spacetime dimensions of the form $A\partial A$. The function
$\Pi_{5}(p,q)$ accounts for the finite frequency, finite momentum response of the system. It is
generated in perturbation theory from the Feynman diagrams shown in Fig. \ref{FD} and can be
regarded as arising from the antisymmetric part of the tensor:
\begin{eqnarray}\label{3leggedsup}
\Pi^{\mu\nu\rho}_{5}(p,q)=-ie^2\int \dfrac{d^5 k}{(2\pi)^5}\mathrm{Tr}\left[G_{k-p}
\Gamma^{\mu}G_{k}\Gamma^{\nu}G_{k-q}\Gamma^{\rho} +G_{k+p-q}
\Gamma^{\nu}G_{k}\Gamma^{\mu}G_{k-q}\Gamma^{\rho}\right].
\end{eqnarray} 
The electronic Green's function is a function of a five-momentum vector $k=(k_{0},\mathbf{k})$ given
by $G(k)=(k_{0}-H(\mathbf{k}))^{-1}$.  
For our model with one massive Dirac fermion in $D=4+1$ dimensions \cite{QHZ08} the low energy
propagator is of the form:
\begin{equation}\label{4+1propagator}
G(k)=\dfrac{k_{0}+\Gamma^{a}k_{a}}{k_{0}^2-\mathbf{k}^2}.
\end{equation} 
Using that the $\Gamma_{a}$ matrices satisfy
\begin{equation}
\mathrm{Tr}\left[ \Gamma_{a}\Gamma_{b}\Gamma_{c}\Gamma_{d}\Gamma_{e}\right]=-4\epsilon_{abcde},
\end{equation} 
one can isolate in \eqref{3leggedsup} the antisymmetric term with five $\Gamma_{a}$ matrices to
obtain, in terms of the Feynman parameters $\alpha,\beta,\gamma$ \cite{DW94}: 
\begin{eqnarray}\label{2ndchern0}
\Pi_{5}^{(a)\mu\nu\rho}(p,q)&=&-16ie^2m\epsilon^{\mu\nu\rho\sigma\tau}p_{\sigma}q_{\tau}\int
\dfrac{d^5
k}{(2\pi)^5}\int^{1}_{0}d\alpha d\beta
d\gamma\hspace{2mm}\dfrac{\delta(1-\alpha-\beta-\gamma)}{\left( k^2-m^2+ p^2\alpha\beta +
q^2\gamma\alpha 
+ (p + q)^2\beta\gamma\right)^{3} },\\
\label{2ndchern}
&=&-\dfrac{e^2m}{8\pi^2}\epsilon^{\mu\nu\rho\sigma\tau}p_{\sigma}q_{\tau}\int^{1}_{0}d\alpha d\beta
d\gamma\hspace{2mm}\dfrac{\delta(1-\alpha-\beta-\gamma)}{\sqrt{m^2-p^2\alpha\beta-q^2\gamma\alpha 
-(p + q)^2\beta\gamma}} \equiv \epsilon^{\mu\nu\rho\sigma\tau}p_{\sigma}q_{\tau}\Pi_{5}(q,p).
\end{eqnarray} 
It is not difficult to check that this definition of $\Pi_{5}(p,q)$ is analogous to that given in
the main text. 
To compute it, it is possible to numerically evaluate the integrals on the Feynman parameters and
find
$\Pi_{5}(q,p)$. \\
As shown in the main text, it is important to keep in mind that in order 
to calculate the finite frequency response to an external time dependent
but spatially uniform electric field, only the external frequency $p_{0}$ is kept finite while all
the rest are sent to zero.  
The integrals in Feynman parameters are analytic and give the logarithmic dependence shown in
\eqref{firstpeak} in
the main text. \\
Consistent with the DC response of a single Dirac fermion \cite{QHZ08}, at $p_{0}=0$ the response is
quantized to $e^2/2h$, also in agreement with the fact that in the lattice model the integrals span
only half of the BZ. The theory recovers the fact that at the boundary of a TI with broken time
reversal symmetry there is a half-integer quantum Hall effect.

\section{Finite chemical potential: $\Pi_{5}(p_{0},\mu)$}

In this appendix we discuss the details of the computation of the response at finite frequency
$p_{0}$ and chemical 
potential $\mu$.
The integral to be computed in this case is defined in
\eqref{2ndchern0} with the replacement $k_{0}\rightarrow k_{0}+\mu$:
\begin{eqnarray}
\Pi_{5}(p,q)&=&-16ie^2m\int \dfrac{d^5 k}{(2\pi)^5}\int^{1}_{0}d\alpha d\beta
d\gamma\hspace{2mm}\dfrac{\delta(1-\alpha-\beta-\gamma)}{\left((k_0+\mu)^2 -\mathbf{k}^2-m^2+
p^2\alpha\beta + q^2\gamma\alpha 
+ (p + q)^2\beta\gamma\right)^{3} }.
\end{eqnarray}
Following the arguments in the main text the relevant case is where all external momenta are zero
except $p_0$. The integral in $k_0$ has two third order poles at $k^{\pm}_0$. The position of the
poles in the complex plane is determined by the relative magnitude of $m^2$, $\mu$, $\alpha$, $p_0$
and $\mathbf{k}^2$. \\
Consider the simple case where $p_0\rightarrow 0$. Following the procedure in Ref.
\onlinecite{CGV10} in this case there is a
pole which is always has a negative imaginary part, no matter what value of $\mathbf{k}$ it has. The
other pole however depending on $\mathbf{k}$ will change semi planes and so for certain values of
$\mathbf{k}$ the integral should be split into two. At this point it is possible to identify several
cases:\\

$|\mu|\leq m$:  In this case there are always both poles in different semiplanes and the integral is
proportional to $\mathrm{sign}(m)$.\\

$|\mu|\geq m$: In this case it is necessary split the integral on $\mathbf{k}$ into two parts. One
from $0$ to $k^{*}$ and the other one from $k^{*}$ to $\infty$ where $k^{*}$ is the value of
$\mathbf{k}$ at which the pole changes semiplane, namely $\sqrt{\mu^2-m^2}$. The first integral
gives zero since both poles are on the same side. The second one is:
\begin{eqnarray}
\Pi_{5}(\mu)&=&16e^2m\int\dfrac{dk_{0}}{2\pi}\int^{\infty}_{k^*}k^3\dfrac{dk}{(2\pi)^4}2\pi^2\hspace
{2mm}\dfrac{1}{\left((k_0+\mu)^2 -k^2-m^2\right)^3}.
\end{eqnarray}
We calculate first the $k_{0}$ integral with the residue theorem. It is a third order pole so closing the
contour from above and restoring $\hbar$ we find:
\begin{eqnarray}
\Pi_{5}(\mu)&=&\dfrac{1}{2}\dfrac{1}{8\pi^2}\left[-\dfrac{3m}{|\mu |}+\dfrac{m^3}{|\mu |^3}
\right]\dfrac{e^2}{\hbar}.
\end{eqnarray} 
Since we need $\sigma(\mu)\equiv 2\pi\Pi_{5}(\mu)$ we finally obtain \eqref{zerofreq}:
\begin{eqnarray}
\sigma(\mu)=2\pi\Pi_{5}(\mu)
&=&\dfrac{1}{4}\left[-\dfrac{3m}{|\mu |}+\dfrac{m^3}{|\mu |^3} \right]\dfrac{e^2}{h},
\end{eqnarray}
which reduces to the familiar $\frac{1}{2}\frac{e^2}{h}\mathrm{sign}(m)$ contribution of a $2+1$
massive Dirac fermion when $m=\mu$, and has an extra term $\dfrac{m^3}{|\mu |^3}$ compared to the
$D=2+1$ case \cite{CGV10}.\\
For finite  $p_{0}$ and $|\mu|\geq m$ one can generalize the same
arguments and find that for $p_{0}<<2m$ we have
\begin{eqnarray}
\sigma(p_{0}<2m,\mu)&=&\dfrac{1}{4}\left[ -\dfrac{3m}{|\mu |}+\dfrac{m^3}{|\mu |^3}
-\dfrac{mp_{0}^2}{6|\mu |^3}\right]\dfrac{e^2}{h}.
\end{eqnarray}

\end{widetext}

\end{document}